\documentclass[12pt]{article}

\usepackage{bm}
\usepackage{cite}
\usepackage{mathrsfs}
\usepackage{slashed}
\usepackage{graphicx}
\usepackage{hyperref}
\usepackage{verbatim}
\usepackage{color}
\usepackage{enumerate}
\usepackage{authblk}
\usepackage[hang,flushmargin]{footmisc}
\usepackage{lipsum}
\usepackage[font=footnotesize]{caption}
\usepackage{soul}
\usepackage[utf8]{} 
\usepackage{empheq}
\usepackage{amssymb,amsmath,amsfonts}
\usepackage[makeroom]{cancel}
\usepackage[normalem]{ulem}
\usepackage{hyperref}
\usepackage{color}

\numberwithin{equation}{section}

\definecolor{blue-violet}{rgb}{0.54, 0.17, 0.89}
\definecolor{PineGreen}{cmyk}{0.92, 0, 0.59, 0.25}
\definecolor{OliveGreen}{cmyk}{0.64, 0, 0.95, 0.40}
\definecolor{RawSienna}{cmyk}{0, 0.72, 1, 0.45}
\definecolor{Gray}{cmyk}{0, 0, 0, 0.50}
\definecolor{MidnightBlue}{cmyk}{0.98, 0.13, 0, 0.43}
\definecolor{Orange}{cmyk}{0, 0.61, 0.87, 0}
\definecolor{LimeGreen}{cmyk}{0.50, 0, 1, 0}
\definecolor{Green}{cmyk}{1, 0, 1, 0}

\renewcommand{\d}{\partial}

\renewcommand{\tilde}{\widetilde}

\usepackage{mathrsfs}

\def\cA{\mathcal{A}}

\def\cD{\mathcal{D}}

\def\cF{\mathcal{F}}

\def\cH{\mathcal{H}}

\def\cJ{\mathcal{J}}

\def\cL{\mathcal{L}}

\def\cN{\mathcal{N}}
\def\cO{\mathcal{O}}

\def\cU{\mathcal{U}}

\def\cX{\mathcal{X}}

\def\cZ{\mathcal{Z}}

\def\dA{\mathscr{A}}

\def\dD{\mathscr{D}}

\def\dN{\mathscr{N}}

\def\be{\begin{eqnarray}}
\def\ee{\end{eqnarray}}
\def\beann{\begin{eqnarray*}}
\def\eeann{\end{eqnarray*}}
\def\beq{\begin{equation}}
\def\eeq{\end{equation}}
\def\ba{\begin{array}}
\def\ea{\end{array}}
\def\ben{\begin{enumerate}}
\def\een{\end{enumerate}}
\def\bea{\begin{eqnarray}}
\def\eea{\end{eqnarray}}

\begin{document}

\title{\vspace{-70pt} \Large{\sc KdV conformal symmetry
breaking \\ in nearly
AdS$_{2}$}\vspace{10pt}}
\author[a]{\normalsize{Marcela C\'ardenas}\footnote{\href{mailto:marcela.cardenasl@uss.cl}{marcela.cardenasl@uss.cl}}}

\affil[a]{\footnotesize\textit{ Facultad de Ingenier\'ia, Arquitectura y Dise\~{n}o, Universidad San Sebasti\'an,\newline
 Bellavista 7, Recoleta, Santiago, Chile.}}

\date{}

\maketitle

\begin{abstract}
We study the gauge theory formulation of Jackiw-Teitelboim gravity and propose Korteweg-de Vries asymptotic conditions that generalize the asymptotic dynamics of the theory. They permit to construct an enlarged set of boundary actions formed by higher order generalizations of the Schwarzian derivative that contain the Schwarzian as lower term in a tower of SL$(2,\mathbb{R})$ invariants. They are extracted from the KdV Hamiltonians and can be obtained recursively. As a result, the conformal symmetry breaking observed in nearly AdS$_{2}$ is characterized by a much larger set of dynamical modes associated to the stationary KdV hierarchy.
We study quantum perturbation theory for the generalized Schwarzian action including the symplectic measure and compute the one-loop correction to the partition function. We find that despite the non-linear nature of the higher-Schwarzian contribution, it acquires a manageable expression that renders a curious leading temperature dependence on the entropy  $S=\#T^{a}$ for $a$ an odd integer.

\end{abstract}

\newpage
\tableofcontents

\section{Introduction}

The construction of dualities in gauge theories provides natural realizations of the holographic conjecture, interchanging mechanisms in both sectors as a communicating vase. In this work, we are motivated by a lower dimensional holographic realization, where a gravitational model in 1+1 dimensions, the so-called Jackiw-Teitelboim (JT) theory \cite{Jackiw:1984je,Teitelboim:1983ux}, has been connected to a 0+1 quantum mechanical system of $N$ Majorana fermions with random interactions, namely, the Sachdev-Ye-Kitaev (SYK) model \cite{Kitaev:2015, Sachdev_1993, Sachdev:2010um}. The origin of the duality is based on the observation that both models exhibit the same kind of symmetry breaking, which is also ruled by the same effective action, the Schwarzian action \cite{Engelsoy:2016xyb, Jensen:2016pah, Maldacena:2016upp}. From the side of the SYK model, there is an emergent reparametrization symmetry on the Euclidean time $y\rightarrow f(y)$, that is spontaneously broken by the vacuum, and down to SL$(2,\mathbb{R})$ in the IR limit. The Schwarzian action represents in this case a low-temperature effective description that characterizes the breaking from $\text{Diff}(S^{1})$ symmetry to SL$(2,\mathbb{R})$ \cite{Kitaev:2017awl}. On the side of dilaton gravity in $1+1$ dimensions, constructing a well-defined action principle leads to a boundary action that turns out to be invariant under Virasoro symmetry in time reparametrizations, but spontaneous and explicitly
broken to SL$(2,\mathbb{R})$. 
In the recent years, the relationship between gravity and SYK has motivated to deepen into both models and find new boundary conditions for AdS$_{2}$ \cite{Grumiller:2015vaa, Grumiller:2017qao,Godet:2020xpk} enriching the previous learning on the topic \cite{hotta1998asymptotic,Cadoni_1999, Navarro-Salas:1999zer}. Also, to extend to supersymmetric extensions \cite{Fu:2016vas, Murugan:2017, Bulycheva:2018qcp, Biggs:2023mfn} or other generalizations like complex SYK models \cite{Bulycheva_2017, Chaturvedi:2018uov, Gu_2020}.

As for the guiding line of this article, we place ourselves in one side of the duality, the gravitational one, and look for generalizations of the asymptotic dynamics of JT Gravity in AdS$_{2}$. It is believed that corrections away from AdS$_{2}$ have a universal form in terms of the Schwarzian derivative \cite{Almheiri:2014cka, Maldacena:2016upp}. In this context, this article explores if there is a manner to realize the exact same symmetry breaking but through a different boundary theory. In order to do so, we look for new boundary conditions on the dilaton $\cX$ and gravitational fields and explore if this entails consequences in the asymptotic dynamics and the path integral computation. We find that a model with these features exists and endow the boundary theory with infinite set of commuting Hamiltonians associated with the Korteweg de Vries (KdV) hierarchy. We find that there is in fact an infinite set of possible boundary values for $\cX$, where the resulting action is still SL$(2,\mathbb{R})$ invariant, and can actually be constructed from an infinite non-linear (but solvable as they arise from a solvable integrable system) set of M\"obius invariants. When written in terms of the reparametrization mode $f(y)$, they can be understood as higher-Schwarzian actions, which are higher order generalizations of the Schwarzian derivative known due to Aharonov, Bertilsson, and Schippers \cite{aharonov1969necessary,Tamanoi1996HigherSO,kim2009some}, and more recently studied in \cite{Galajinsky:2023btq, Krivonos:2024jpo}, where they have tried to unravel their relationship with symmetries transformations. Following this scheme, in our of work the Schwarzian action on the reparametrization mode $f(y)$ results to be the lowest term in a tower of invariants. In this respect, this article proposes a more general set of boundary conditions, as it includes the previous result, and offers and enlarged set of possible fluctuating behaviours of the  gravitational field at the boundary. Indeed, it connects it with the dynamics of the stationary KdV hierarchy. 
We found a Hamiltonian vector that depends locally on the fields and ensures the invariance of the symplectic two-form under the generalized KdV symmetries. We reconstruct the boundary theory from the symplectic two-form that serves us to define a measure in the path integral and recover the Hamiltonians obtained from the surface term associated to the BF theory. We study the holonomies associated to the gravitational gauge field and use them to characterize our boundary conditions from two manners: their trace is an invariant element of the group that is conserved and permits to find local conserved charges. Also, demanding them to be trivial along the thermal cycle fixes the black hole temperature.
We compute the one-loop correction to the partition function and show that despite the non-linear nature of the generalized Schwarzian terms, it renders a manageable equation that leads to a power law temperature-dependence of the entropy. At this point, we find similarities with other SYK and matrix models  where they have found a low energy thermodynamics characterized by a curious scaling of the entropy in $T$ \cite{Lin:2013jra,Biggs:2023mfn}.

This work also contribute to extend the perspective of other interesting works on 2-dimensional dilaton gravity and KdV that have considered a genus expansion and the properties of integrability to compute the partition function \cite{Okuyama:2019xbv,Blommaert:2022lbh}. The relationship between conformal field theories and KdV has been also explored in \cite{Sasaki:1987mm,Kupershmidt:1989bf,Bazhanov:1994ft,Asrat:2020jsh}.
 
This article is organized as follows:
In section \ref{Bf theoyr} we define a gauge theory formulation of Dilaton gravity by means of the BF model. In section \ref{sec:3}, we present our set of
boundary conditions, study the associated asymptotic symmetries and field equations. In section \ref{sec:4}, we solve the integrability problem of the Hamiltonian using the proposed asymptotic conditions and construct an infinite tower of KdV Hamiltonias as a boundary action for JT gravity. In section \ref{sec:5}, we present the first order formalism and show the boundary conditions in terms of the metric field. We also present an improper gauge transformation for AdS$_2$ that permits to reconstruct a generalized nearly AdS$_2$ spacetime, that contains the KdV perturbations. In section \ref{sym-bt} we rebuild the boundary theory using the symplectic structure of KdV. In section \ref{sec:7}, we study the holonomy group element. In section \ref{Partition function}, we compute the perturbed partition function up to quadratic contributions and obtain the entropy.

\section{Dilaton gravity as a BF-theory} \label{Bf theoyr}
We use the gauge formulation of Dilaton gravity as a non-abelian model, 
\begin{equation}\label{action}
I[\cX,\cA]=\frac{k}{2\pi}\int_{\mathcal{M}} {\rm tr}\left[\cX \cF \right] + I_{\rm bndy}
\end{equation}
being $\cA$, the one-form connection associated with the field strength $\cF$ and $k$ a coupling constant. The dilaton field $\cX$ is an algebra-valued scalar. The topology of $\mathcal{M}$ is chosen to be $S^1$, where the radial coordinate ranges as $0\leq\rho<\infty$. Furthermore, the Euclidean time $y$ is identified as $y \sim y +\beta$ with $\beta$ being the inverse of the temperature. Additionally, $I_{\rm bndy}$ stands for a suitable boundary term implemented to have a well-defined action principle for some given non-trivial boundary conditions at $\partial \mathcal{M}$, which is the surface of infinite radius.\\
The relevant gauge group is SL$(2,\mathbb{R})$, for which we choose a matrix representation for its generators,
\begin{equation}
L_{-1}=\begin{pmatrix}0 & 0\\
1 & 0
\end{pmatrix}\quad,\quad L_{0}=\begin{pmatrix}-\frac{1}{2} & 0\\
0 & \frac{1}{2}
\end{pmatrix}\quad,\quad L_{1}=\begin{pmatrix}0 & -1\\
0 & 0
\end{pmatrix}\;.\label{PTT-sl(2,R)-MR}
\end{equation}
The algebra of $L_{n}$ satisfy the commutation relation
$\left[L_{n},L_{m}\right]=\left(n-m\right)L_{n+m}$, and the only non-zero components of the invariant bilinear form are given by $\langle L_{0}, L_{0}\rangle=1/2$ and $\langle L_{1},L_{-1}\rangle=-1$. 

The action is invariant under the gauge transformations, 
\begin{equation}\label{gauge tranform}
\delta_{\Lambda} \cA =d\Lambda + \left[\cA,\Lambda \right],\qquad \qquad \delta_{\Lambda} \cX =[\cX,\Lambda],
\end{equation}
where $\Lambda$ is a Lie-valued gauge parameter. The equations of motion are given by
\begin{equation} \label{eom}
\mathcal{F} =d\cA+\cA\wedge \cA=0,\qquad \qquad d \cX + \left[\cA,\cX\right]=0.
\end{equation}

\section{KdV boundary conditions}  
\label{sec:3}
In this section, we build generalized KdV boundary conditions for JT gravity following the lines of \cite{Perez:2016vqo,Gonzalez:2018jgp,Cardenas:2021vwo}. We take advantage of the role  played by the spectral parameter $\lambda$ in integrable systems, so that we can write down an iterative expansion for the dilaton field at infinity. The latter will enable us to obtain a boundary Hamiltonian consistent with the symmetries observed in the KdV hierarchy.

\subsection{Asymptotic symmetries}
\label{asyIS}
In order to analyse the set of configurations consistent with the field equations \eqref{gauge tranform} and the well-possedness of the action principle \eqref{Ib}, we write $\cA$ and $\cX$ in terms of auxiliary fields $a=a_y dy$ and the adjoint scalar $x$, 
\be \label{radial gaugetrans}
\cA=b^{-1}(\rho)(d +a)b(\rho)\, \quad \cX=b^{-1}(\rho) \, x(y) \, b(\rho)\,,
\ee
with a fixed group element $b=\exp[\rho L_0]$ capturing the radial dependence. Inspired by boundary conditions devised in  three-dimensional gravity \cite{Perez:2016vqo,Cardenas:2021vwo}, we suggest
\begin{equation}
\begin{split}
a_{y}(\cL;\lambda)& =- \frac{2\pi}{k}\cL(y)L_{-1}+L_{1}-2\lambda L_{0},\\
x(\mu,\cL;\lambda)& = \mu L_{1}- \left(\mu'+ 2\lambda  \mu \right) L_{0}+\left(\frac{\mu''}{2}+\lambda  \mu'-\frac{2 \pi  \mu \mathcal{L}}{k} \right)L_{-1}\,,
\end{split}
\label{eq:kdvbc}
\end{equation}
where $\lambda$ is a constant without variation, as proposed in \cite{Cardenas:2021vwo}. The profile of the auxiliary fields 
can be regarded as an extension of the Brown-Henneaux boundary conditions, nonetheless by including this parameter in the boundary conditions, permits to provide a self-cointained mechanism to construct a well-defined variational principle connecting the boundary dynamics with the KdV hierarchy, as it shown thoroughly in the appendix \ref{Appendix }. Once our boundary dynamics is completely characterized, $\lambda$ can be set to zero. 
Gauge transformations preserving \eqref{eq:kdvbc} are those mapping configurations into themselves,
\be
\delta_{\Lambda} a=\cO(a)\,,\quad \delta_{\Lambda} x=\cO(x)\,.
\ee
These transformations can be solved for the gauge parameter $\Lambda$ modulo trivial transformations.\footnote{A trivial transformation is generated by an antisymmetric tensor $\chi_{\mu\nu}=-\chi_{\nu\mu}$, where  $\delta_{\chi} A_{\mu}=\chi_{\mu\nu}\frac{\delta I}{\delta A_{\nu}}$. This transformation acts trivially in the action variation as follows,
\begin{equation}
  \frac{\delta I}{\delta A_{\mu}}\delta_{\chi} A_{\mu}=  \frac{\delta I}{\delta A_{\mu}}\frac{\delta I}{\delta A_{\nu}}\chi_{\mu\nu}=0.
\end{equation}
}
The solution that respects the asymptotic conditions \eqref{eq:kdvbc} takes the functional form of the dilaton field with parameter $\varepsilon$,
\be
\label{dilatongauge}
\Lambda=x(\varepsilon,\cL;\lambda), 
\ee
and the asymptotic symmetry transformations for the relevant fields $\cL$ and $\mu$ reduce to 
 \begin{equation}\label{symmetry trasnformations}
 \begin{split}
    \delta_\varepsilon \cL&= \varepsilon\cL'+2\varepsilon'\cL-\frac{k}{4\pi}\varepsilon'''+\lambda^2 \frac{k}{\pi}\varepsilon'\,,\\
\delta_\varepsilon \mu&=  \varepsilon \mu'-\varepsilon'\mu \,. 
    \end{split}
\end{equation} 
One can readily see that $\mu$ transforms as a vector at the boundary, while $\bar{\cL}=\cL-\frac{k}{2\pi}\lambda^2$ corresponds to a spin-two density. In the present work, we keep the explicit dependence on $\lambda$, as it will serve as a parameter to recursively solve our boundary conditions in the next section. 

In sum, the residual symmetries associated to $a_y$ and $x$ are diffeomorphisms on the circle generated by a vector field $\varepsilon$.

\subsection{Field equations}

On-shell, the  vanishing of the curvature $\cF$ is trivially fulfilled by \eqref{eq:kdvbc}. Thus, it results that the dynamics of the system is only dictated by the dilaton equation
 \begin{equation} \label{eomdilaton}
\mu \cL' +2\cL \mu' - \frac{k}{4\pi} \mu'''+\frac{k}{\pi}\lambda^2 \mu' = 0,
\end{equation}
that corresponds to the symmetry transformations leaving invariant the function $\cL$. The function $\mu$ defining the adjoint scalar $x$ admits an expansion in  powers of $\lambda^2$,
\begin{equation}\label{expansion}
\mu=\sum_{n=1}^{N+1}\mu_{n}\lambda^{2(N+1-n)},
\end{equation}
where $N\in\mathbb{Z}^{0+}$. This method introduces a systematic way to solve the dilaton equation; i.e.  recursively and taking powers of $\lambda^2$. The first equation that can be extracted is the term of order $\lambda^0$, which yields, 
\begin{equation}\label{eomKDV}
\mu_{N+1} \cL' +2\cL \mu_{N+1}' - \frac{k}{4\pi} \mu_{N+1}'''=0 \,.
\end{equation}
 This is the field equation for the dilaton. The remaining equations associated to higher orders of $\lambda$  can be presented in a compact form, employing the differential operator
\be \label{operator1}
\mathcal{D}= \cL' +2\cL \partial_{y} - \frac{k}{4\pi} \partial^{3}_{y}\,,
\ee
and they are given by
\begin{equation}\label{recurrence}
 \mu'_{n+1}=-\frac{\pi}{k}\mathcal{D}\mu_{n}\,,\quad\text{for}\quad 1\leq n \leq N
\end{equation}
with the condition $\mu'_{1}=0$. 
This recursion relation produces a large amount of distinct dynamical modes for the boundary dilaton, controlled by the arbitrary integer $N$. The explicit construction of the generalized dynamics is described in appendix \ref{Appendix }. This formula \eqref{recurrence} forms a basis for the phase space, and permits to construct an infinite tower of commuting Hamiltonians. In fact, one can show that $\mu_n$ can be expressed in terms of a functional density $\cH_n$, so that 
\be
\label{clave}
\mu_n=g_n \,\frac{\delta \cH_n}{\delta \cL}\,,
\ee
where $g_{n}$ is a fixed coupling constant with dimension of $[{\rm length}]^{2n-1}$. It can be regarded as the value of the dilaton field when $\cL$ acquires a constant value.

\section{Hamiltonian} \label{sec:4}
\subsection{Action principle}
\label{BoundaryAction}

Once the field equations \eqref{eom} hold, the infinitesimal variation of the action principle vanishes  after demanding 
\begin{equation} \label{deltaI}
    \delta I_{\rm bndy}+\frac{k}{2\pi}\int_{\partial \mathcal{M}}\,{\rm tr}\left[\cX \delta \cA\right]=0\,.
\end{equation}
The theory is then fully determined provided the boundary term $I_{\rm bndy}$ is known. This is accomplished after suitable boundary conditions are given for both the dilaton $\cX|_{\partial \mathcal{M}}=x$ and the gauge field $\cA|_{\partial \mathcal{M}}=a$. For this purpose, we evaluate \eqref{deltaI} on the asymptotic value of the fields \eqref{eq:kdvbc}. The boundary term reduces to
\begin{equation}\label{deltaH}
\delta I_{\text{bndy}}= -\int dy \, \mu \delta \cL \,.
\end{equation}
Thanks to the functional relation between the dilaton and the gauge field provided in equation \eqref{clave}, the boundary action is readily integrable. Indeed, 
\begin{equation}\label{Ib}
    \mu_n=g_n \frac{\delta \mathcal{H}_n}{\delta \cL}\qquad\text{implies that}\qquad I_{\rm bndy} =g_n \int_{\partial \mathcal{M}}\,\mathcal{H}_n\, .
\end{equation}
As shown in the appendix \ref{Appendix }, the computation of $\mu_{n}$ implies integrating $n$ times the recurrence relation \eqref{recurrence}. As a consequence, there are $n$ unfixed constants of integration labeling $n$-th Hamiltonian of the KdV hierarchy. In what follows, we consider this freedom so that to obtain each Hamiltonian separately. Working out the recurrence \eqref{recurrence}, the first four Hamiltonian yields
\begin{eqnarray}\label{KdvHamiltonians}
 H_1&=&\frac{2\pi }{k}\int dy\,\cL\, ,\\ 
 H_2&=& -\left(\frac{2\pi }{k}\right)^2\,\int dy \, \cL^2\,,\\
 H_3&=& 2\left(\frac{2\pi }{k}\right)^{3} \,\int dy \, \left(\cL^3+\frac{k}{4\pi}\cL'^2\right)\,,\\
 H_4&=&- 5\left(\frac{2\pi }{k}\right)^{4}\,\int dy \, \left(\cL^4+\frac{k}{\pi}\cL\cL'^2+\frac{k^{2}}{20\pi^{2}}\cL''^{2}\right)\,.
\end{eqnarray}
As it can be seen, the dimension of the operator goes like $[{\rm length}]^{1-2n}$, which provides a dimensionless boundary action principle. The action principle constructed in this section is naturally ${\rm SL}(2,\mathbb{R})$
invariant. This can be noticed from \eqref{deltaH} and using that $\delta_{\varepsilon} \cL=0$, for a transformation $\varepsilon$ that preserves the orbit $\cL_0=-\tfrac{\pi^2}{\beta^{2}}$.

\section{Asymptotic conditions in the metric formulation} \label{sec:5}
In this section, we are interested in recovering our boundary conditions written so far as algebra-valued fields, in terms of the second order fields. Firstly, we introduce the first order formulation of JT-gravity \cite{Fukuyama:1985gg} as a preamble to go the metric formulation and analyse our boundary conditions in that framework. We construct the symmetry transformations as diffeomorphisms and propose an improper change of coordinates that renders the final form of nearly AdS$_{2}$ metric with KdV boundary conditions.

\subsection{First order formulation}
 The action principle is given by, 
\begin{equation}\label{action2}
    I_{JT}=\frac{k}{4\pi}\int_{\mathcal{M}}X^{a}(de_{a}+\epsilon_{ab}\omega e^{b})+\Phi(d\omega+\frac{1}{2}\epsilon_{ab}e^{a}e^{b}), 
\end{equation}
where $a,b=0,1$, $e^{a}$ is the vielbein and $\omega^{ab}=\epsilon^{ab}\omega$ is the spin connection. The spin-zero dilaton field is  $\Phi$, whereas $X^{a}$ are Lagrange multipliers related with the torsionless condition and $\epsilon_{ab}$ is the Levi-Civita tensor.

The first order action \eqref{action2} for JT gravity can be recovered from the BF-formulation of the theory \eqref{action}, by identifying
\begin{equation}\label{1st order fields}
 \mathcal{A}_{I}=(e_{0},e_{1},\omega)\quad\text{and}\quad \cX^{I}=(X^{0},X^{1},\Phi)  
\end{equation}
 for $\{I=a,2\}$. These fields are spanned in terms of the 
$\mathfrak{
so}(2,1)$ algebra,
\begin{equation}
 \left[T_{I},T_{J}\right]=\eta^{KL}\epsilon_{IJK}T_{L}   
\end{equation}
 with the associated invariant bilinear form $\eta_{IJ}=2\langle T_{I},T_{J}\rangle= \operatorname{diag}(1,1,-1)$ and the antisymmetric invariant tensor $\epsilon^{012}=1$, that is related with the one in the action \eqref{action2} taking $\epsilon_{ab}=\epsilon_{ab2}$. 
 
  The metric components in terms of the vielbein can be retrieved using the formula 
\begin{equation}\label{metric-vielbein} 
    g_{\mu\nu}=\eta_{ab}e^{a}_{\mu}e^{b}_{\nu}=2 \rm tr(e_{\mu}e_{\nu}).
\end{equation}
The relevant algebra along this work has been $\mathfrak{sl}(2,\mathbb{R})$, which is isomorphic to $\mathfrak{so}(2,1)$. To compare fields in both basis, we consider the mapping,
\begin{equation}
 T_{2}=\frac{1}{2}\left(L_{-1}+L_{1}\right),\text{\ensuremath{\quad}}T_{0}=\frac{1}{2}\left(L_{-1}-L_{1}\right),\quad T_{1}=L_{0}.   
\end{equation}
The gauge field $\mathcal{A}=\mathcal{A}^{+}L_{1}+\mathcal{A}^{-}L_{-1}+\mathcal{A}^{0}L_{0}$ and the algebra-valued dilaton field $\cX=\cX^{+}L_{1}+\cX^{-}L_{-1}+\cX^{0}L_{0}$ proposed in \eqref{radial gaugetrans} are connected to the $\mathfrak{so}(2,1)$ components of vielbein and the spin connection \eqref{1st order fields} via
\begin{eqnarray}\label{vielbeins}
    e^{0}=\mathcal{A}^{+}-\mathcal{A}^{-},\qquad e^{1}= \mathcal{A}^{0},\qquad \omega=-(\mathcal{A}^{+}+\mathcal{A}^{-}).
\end{eqnarray}
The relations for the Lagrange multipliers and the dilaton field are ruled by the equations,
\begin{eqnarray}\label{Dilaton eq}
    X^{0}=\cX^{+}-\cX^{-},\qquad X^{1}= \cX^{0},\qquad \Phi=\cX^{+}+\cX^{-}.
\end{eqnarray}  
\subsection{Metric formalism} \label{Metric formalism}
Considering  the prescriptions of the previous subsection, we use the identifications \eqref{vielbeins} to construct the vielbiens in terms of our asymptotic conditions presented as a one-form gauge field \eqref{eq:kdvbc}. Then, we employ the formula \eqref{metric-vielbein} so that to pass to a metric form, 
\begin{equation} \label{metric}
 ds^2=\left[\left( e^\rho+\frac{2\pi \cL}{k}e^{-\rho} \right)^2+4\lambda^2\right]dy^2- 4\lambda\, dyd\rho + d\rho^2.
\end{equation}
Here, the radial gauge transformation in \eqref{radial gaugetrans} was chosen as $b(\rho)=e^{(\rho L_0)}$, to make contact with a generalized Fefferman-Graham expansion \cite{Fefferman:2007rka, Grumiller:2017qao}.
Likewise, starting from the components of the algebra-valued dilaton proposed in \eqref{eq:kdvbc}, we use \eqref{Dilaton eq} to construct the dilaton field in the second order formalism, 
\begin{equation} \label{dilaton 2nd}
    \Phi=e^{\rho}\mu+e^{-\rho}\left(\frac{\mu''}{2}+\lambda \mu '-\frac{2\pi\mu\cL}{k}\right).
\end{equation}
We look for the set of diffeomorphisms $\xi^\mu=\left\{\xi^y,\xi^\rho\right\}$ preserving the form of \eqref{metric} and \eqref{dilaton 2nd}, that obey the equations
\begin{equation}\label{geometric symm trasnf}
    \mathcal{L}_{\xi}g_{\mu\nu}=\mathcal{O}(g_{\mu\nu}),\qquad \mathcal{L}_{\xi}\Phi =\xi^{\mu}\partial_{\mu}\Phi =\mathcal{O}(\Phi).
\end{equation}
We find that the components of the asymptotic Killing vectors solving the above are,
\begin{equation}
  \xi^y(y,\rho)=\varepsilon+\frac{k\, \alpha'}{2\left(k\,e^{2\rho}+2\pi \cL \right)}\qquad\text{and}\qquad  \xi^\rho(y,\rho)=\alpha-2\lambda \xi^y,
\end{equation}
where $\varepsilon=\varepsilon(y)$ and $\alpha=2\lambda\varepsilon-\varepsilon'$. We also obtain that the asymptotic symmetry transformations \eqref{geometric symm trasnf} reduce to \eqref{symmetry trasnformations} up to trivial gauge transformations. This shows the consistency between the transformations of the geometric fields with the gauge transformations in \eqref{gauge tranform}, choosing the gauge parameter as $\Lambda=\cX(\varepsilon)$.\\
We propose now a coordinate change from AdS$_{2}$ to a metric of the form \eqref{metric}, that realizes the above infinitesimal symmetry in an exact form. Based on the three-dimensional case  \cite{Roberts:2012aq}, let us consider AdS$_{2}$ in radial coordinates with Lorentzian time $T$,
\begin{equation}
    ds^{2} = -R^{2} d T^2+R^{-2} d R^2,
\end{equation}
and perform the change of coordinates
\begin{equation}
    T=g(t)+\frac{2\left(\partial_t g\right)^2\,\partial^{2}_t g}{4 e^{2\bar{\rho}}\,g^2- \partial^{2}_t g},\qquad
    R=\frac{e^{\bar{\rho}}}{\partial_t g}-\frac{\partial^{2}_t g}{4 e^{\bar{\rho}}\,\left(\partial_t g\right)^3},
\end{equation}
where we additionally take $t\rightarrow-it$, in order to go to the Euclidean signature. This coordinate change renders the metric,
\begin{equation}
    ds^2=\left(e^{\bar{\rho}}-\frac{\left\{g,t\right\}}{2}e^{-\bar{\rho}}\right)^2\,dt^2+d\bar{\rho}^2.
\end{equation}
Here $\left\{g,t\right\}$ stands for the Schwarzian derivative, which is defined as 
\begin{equation}
  \left\{g,t\right\}=\frac{g'''}{g'}- \frac{3}{2}\left(\frac{g''}{g'}\right)^{2},
\end{equation}
which is invariant under the action of the SL$(2,\mathbb{R})$ transformation
\be \label{Shwar trans}
g(y)\to \tilde{g}(y)=\frac{ag(y)+b}{cg(y)+d}\,\quad\text{with}\quad ad-bc=1.
\ee

We perform now a second coordinate change. We take $t=h(y)$ and the additional redefinition of the radial coordinate $e^{\rho}=h ' e^{\bar{\rho}}$. This can be seen as a double reparametrization on the temporal coordinate $T\rightarrow g\circ h$ as $\rho\rightarrow\infty$. Applying the new redefinition of the coordinates on the above metric, we obtain,
\begin{equation}
\label{metrictrans}
    ds^2=\left[\left(e^{\rho}-\frac{1}{2} (h')^2 \{g\circ h,h\} e^{-\rho}\right)^2+\left(\frac{h''}{h'}\right)^2\right]\,dy^2+\frac{2h''}{h'}dy d\rho+d\rho^2\,.
\end{equation}
This spacetime can be linked to the one associated to our boundary conditions \eqref{metric}, by choosing, 
\begin{equation}
    h(y)=e^{-2\lambda\,y}\qquad\text{and identifying}\qquad \cL=-\frac{k}{4\pi}(h')^2\left\{g\circ h,h\right\}.
\end{equation}
The transformation law of the Schwarzian derivative under function composition
\begin{equation}
 (h')^{2}\left\{g\circ h,h\right\} =\left\{g\circ h,y\right\}-\left\{h(y),y\right\},
\end{equation}
permits to write explicitly that 
\begin{equation}\label{L en schwarzian}
   \cL=-\frac{k}{4\pi}\left(\left\{g\circ h,y\right\}+2\lambda^2\right).
\end{equation}
As a final step in the characterization of the metric functions, we define 
\begin{equation}
g\circ h=\text{Tan}\left[\sqrt{-\mathcal{L}_{0}-\lambda^2}f(y)\right],
\end{equation}
so that to relate $\cL$, as defined in \eqref{L en schwarzian}, with the symmetries of the little group associated to regular black hole solutions \eqref{metric}. In terms of the reparametrization mode $f$, we find
\begin{equation}\label{rep}
\cL=\frac{k}{2\pi}\left((\mathcal{L}_{0}+\lambda^2)f'^{2}-\frac{1}{2}\left\{f,y\right\}-\lambda^{2}\right),
\end{equation}
where $\cL_{0}$ is a constant linked to the constant representative solution.

\section{Symplectic structure of the boundary theory} \label{sym-bt}
An alternative manner to construct the boundary action is through the symplectic structure of the KdV dynamics. A manifold $\mathcal{M}$ is called symplectic when it is endowed with a symplectic two-form $\omega$ over the phase space. We consider the notation $\iota_{V}$ to denote the interior product with respect to a vector field $V$, and also the exterior derivative $d$, which operates on the associated phase space. 
Given a vector field $V$, the invariant action of $V$ over $\omega$ imply, 
\begin{equation} \label{derivadalie}
    (d\,\iota_{V}+\iota_{V}\,d)\omega=0,
\end{equation}
which is the vanishing Lie derivative of $V$ acting on $\omega$. When $\omega$ is exact, then condition of invariance on $\omega$ \eqref{derivadalie} reduce to $d(\,\iota_{V}\omega)=0$, or equivalently, to
\begin{equation}\label{invariancecondition}
    \iota_{V}\omega=dH.
\end{equation}
The previous equation establishes that finding a Hamiltonian function $H$ is equivalent to proving the invariance of $\omega$. In this case, it is said that the action of $V$ on the phase space $M$ is symplectic. 
We use the symplectic two-form derived from the orbit of a particular coadjoint vector under the action of the Virasoro group of\cite{Witten:1987ty, Bakas:1988bq, Alekseev:1988ce}, 
\begin{equation}\label{symplectic virasoro}
 \omega=-\frac{k}{2\pi}\int dy \left[\cL_{0} df\,df'+\frac{1}{4}\frac{df'}{f'}\left(\frac{df'}{f'}\right)'\right]\,, 
\end{equation}
which is the finite version of the symplectic form that will be derived in section \ref{Partition function}. 

The symplectic form is invariant under Diff $(S^1)$, whose transformation is infinitesimally given by $\delta y=\epsilon(y)$, and acts on the field as
\begin{equation}\label{transformation}
  \delta f(y)=  \epsilon(y)f'.
\end{equation}
The resulting Hamiltonian, obtained from applying \eqref{invariancecondition} is given by
\begin{equation}
    H=\frac{k}{2\pi}\int dy \left[\cL_{0}f'^2-\frac{1}{2}\{f,y\}\right].
\end{equation}
To generalize these results and recover the boundary theory computed from the surface term of the BF-theory, we propose a field-dependent Hamiltonian vector
$\delta y=\mu(f(y))$, which acts on the field as
\begin{equation}\label{transformation2}
  \delta f(y)=  \mu(f(y))f'.
\end{equation}
Assuming the dependence of $\mu$ on the fields is set by KdV recurrence relation \eqref{recurrence}, one can construct a Hamiltonian vector $\tilde{V}_{k}=\mu_{k} V$.

We prove now that the action of $\tilde{V}_{k}$ on $\omega$ is symplectic. We start considering the fundamental exterior product   $\iota_{\tilde V_{k}} df= \mu_{k} f' $, and that  $\iota_{\tilde V_{k}} $ acts on  \eqref{symplectic virasoro}, such that

\begin{equation}
 \iota_{\tilde V_{k}}\omega=-\frac{k}{2\pi}\int dy \left[\cL_{0} \iota_{\tilde V_{k}}(df\,df')+\frac{1}{4}\iota_{\tilde V_{k}}\left(\frac{df'}{f'}\left(\frac{df'}{f'}\right)'\right)\right]\,.  
\end{equation}
The first term in the integral can be worked out easily to show that it can be factorized as $\mu_{k}d\left[\cL_{0}(f')^{2}\right]$ up to some boundary terms. The second term needs a more careful treatment and various integration by parts, but the final result ends up being the compact expression
\begin{equation}
 \iota_{\tilde V_{k}}\omega=-\frac{k}{2\pi}\int dy \mu_{k}d\left[\cL_{0}(f')^{2}-\frac{1}{2}\left\{f,y \right\}\right]\,.  
\end{equation}
Provided the functional form of $\mu_{k}$ is the one proposed in \eqref{clave}, we have that
\begin{equation}
    \iota_{\tilde{V}_{k}}\omega=d H_{k},
\end{equation}
where $H_{k}$ is the $k$-th Hamiltonian of the KdV hierarchy \eqref{KdvHamiltonians}. Here we have also consider the reparametrization found in \eqref{rep} with  $\lambda=0$. 

\section{Holonomy: Conserved quantities and thermodynamics} \label{sec:7}

In this section, we take as a fundamental object the holonomy associated to a gauge field $A$,
\begin{equation} \label{holonomy}
\text{Hol}[A]=\mathcal{P}{\rm exp}\left(\int_{\gamma} A \right),
\end{equation}
which is a path-ordered parallel transport exponential defined along a curve $\gamma$. Holonomies will be a central object to characterize our boundary conditions from two viewpoints. On the one hand, the trace of $\text{Hol}[A]$ is an invariant element of the group and permits to construct conserved quantities.  It will provide and alternative manner to recover the KdV Hamiltonians previously obtained from the boundary action \eqref{KdvHamiltonians}.
On the other, demanding the holonomy to be trivial along the thermal cycle fixes $\beta$ to the inverse of the black hole temperature.

\subsection{Conserved quantities}
We evaluate \eqref{holonomy} for the asymptotic conditions defined in \eqref{eq:kdvbc} on a closed path,
\begin{equation} \label{holonomy1}
\text{Hol}[a_{y}(y;\lambda)]=\mathcal{P}{\rm exp}\left(\oint a_y (y;\lambda)  dy\right).
\end{equation}
The above is a gauge group element evaluated on the phase space and independent of the radial coordinate by construction. As a consequence, it is always satisfied that $\partial_{\rho}\text{Hol}[a_{y}(y;\lambda)]=0$, which despite of being a trivial conservation law, permits to construct non-trivial conserved quantities along the radial direction. They are in fact, an infinite number after expanding the trace of \eqref{holonomy1} on powers of the spectral parameter.\footnote{For 2D integrable systems with a Lax Pair structure, it is possible to show that the trace of the holonomy on a closed loop leads to infinite non-local integrals of motion (see e.g. \cite{Babelon:2003qtg}).}

To simplify the computation of the holonomy, we look for a gauge transformation $g$ that diagonalizes our boundary conditions \eqref{eq:kdvbc}. For the chosen matrix representation of the $\mathfrak{sl}(2,\mathbb{R})$ algebra, this implies finding a gauge connection of the form $\mathfrak{a}_y=\cJ(y) L_0$, where
\begin{equation} \label{gaugetransf2}
  \mathfrak{a}_y=g^{-1} a_y g+ g^{-1} \partial_y g.
\end{equation}
In general, the holonomy evaluated in two points $(y,y')$ transforms under gauge transformations as 
\begin{equation}
\text{Hol}[\mathfrak{a}_y]=g^{-1}(y)\text{Hol}[a_y]g(y').
\end{equation}
Using the above equation, evaluating it on closed loop where $y'=y+2\pi$, and considering $g(y)=g(y+2\pi)$, one obtains
\begin{equation}
{\rm tr}(\text{Hol}[\mathfrak{a}_y])={\rm tr}(\text{Hol}[a_y]).
\end{equation}
Then, the trace of the holonomy is invariant under periodic gauge transformations. 

We calculate now the precise group element in \eqref{gaugetransf2} that permits to write the connection in a diagonal form. Generically, it can be expressed as $g=e^{f\,L_{-1}}e^{h\,L_1}e^{p\,L_0}$, for some periodic functions $p(y),h(y),f(y)$. One can check that the diagonalization condition \eqref{gaugetransf2} reduces to
\begin{equation} \label{ecuacion-offdiagonal}
  \cL=\frac{k}{2 \pi}\left(-2\lambda\,f+f^2-f'\right),\qquad f=\frac{1+2\lambda\,h-h'}{2h},  
\end{equation}
which kills the off-diagonal sector. These two conditions combined take the form 
\begin{equation} \label{ecuacion-diagonal}
    \frac{8\pi}{k}\cL+4\lambda^2=\frac{1}{h^2}-\frac{h'^2}{h^2}+2\frac{h''}{h}.
\end{equation}
Additionally, the diagonal part of \eqref{gaugetransf2} leads to the equation
\begin{eqnarray}\label{eqdeJ}
    \cJ=\frac{1-h'-h\,p'}{h}.
\end{eqnarray}
One would like to relate the field $\cJ$ in $\mathfrak{a}_y$ with the original field $\cL$. This is possible mixing the equations \eqref{ecuacion-diagonal} and \eqref{eqdeJ} and choosing $p=\frac{-2h'}{h}$, which gives the final relation
\begin{equation} \label{expan J}
    \cJ^2+\cJ'=\frac{2\pi}{k}\cL+\lambda^2.
\end{equation}
The above is a Miura transformation that can be understood as Ricatti equation for $\cJ$, and permits to construct a recursion relation that recovers the KdV Hamiltonian densities proposed in the previous section. In general, this equation can be solved by considering the expansion,
\begin{equation} \label{expansionJ}
    \cJ=\lambda+\sum_{l=1}^{\infty} \cJ_{l}(2\lambda)^{-l},
\end{equation}
and solving \eqref{expan J} order by order in $\lambda$. From this procedure, it is found that $\cJ_{0}=0$, $\cJ_{1}=2\pi\,\cL/k$, and 
\begin{equation} \label{Gerlfand2}
    \cJ_{l+1}=-\cJ_{l}'-\sum_{m=1}^{l-1}\cJ_{l-m}\cJ_{m},\quad l\geq1.
\end{equation}
This expression offers an alternative recursion to generate the densities associated to the KdV Hamiltonians \footnote{Indeed, $\cJ_{2}=-\frac{2\pi}{k}\cL'$, $\cJ_{3}=-\frac{4\pi}{k^{2}}(\cL^{2}-\frac{k}{2\pi}\cL'')$, $\cJ_{4}=-\frac{16\pi^{2}}{k^{2}}(-\cL'''+\frac{k}{8\pi}\cL\cL')$,$\cdots$, where the terms with even subindices lead to total derivatives.}. In this case, they are given by 
\be
H_{n}=\int \cJ_{2n-1} dy\,=\int \cH_{n}dy\quad\text{with}\quad n\geq1.
\ee
Considering the reparametrization proposed in 
\eqref{L en schwarzian}, the recurrence \eqref{Gerlfand2} provides an infinite set of densities that depend on powers and derivatives of the Schwarzian. By consequence, they keep realizing the SL$(2,\mathbb{R})$ symmetry via the reparametrization mode that transforms according to \eqref{Shwar trans}, nonetheless they introduce new infinite possible non-linear boundary actions fulfilling the invariance.

\subsection{Regularity conditions}
 The regularity condition fixes the value of the period of the thermal cycle $y \sim y+ \beta$, after imposing the holonomy to be equivalent to the center of the group, 
\begin{equation} \label{trivial holonomy}
\text{Hol}[\mathfrak{a}_{y}]=\mathcal{P}{\rm exp}\left(\oint \mathfrak{a}_y dy\right)=-\mathbb{I}.
\end{equation}
As we are only interested in configurations subject to thermal equilibrium, so that \eqref{expan J} reduces to
\begin{equation} \label{expanJ2}
    \cJ=\pm\sqrt{\cL_{0}+\lambda^2},
\end{equation}
where $\cL_{0}$ is the zero mode appearing in the reparametrization \eqref{rep}. At this point, we remind the nature of $\lambda$, which acts as a generating parameter of the generalized asymptotic dynamics, and it was initially considered to be devoid of physical meaning, so we set it equal to zero. Then, 
the trivial holonomy condition \eqref{holonomy} reduces to 
\begin{equation} \label{temperature}
   \cL_{0}=-\frac{n^2\pi^2}{\beta^2}\,,
\end{equation}
with $n$ an arbitrary integer representing $n$-folded covering of the two-dimensional AdS space.

\section{Thermal partition function} \label{Partition function}
The path integral representation of the BF model partition function \eqref{action} has been considered in \cite{Witten:1991we} for compact gauge groups. It can be easily worked out, by noticing that the since the Dilaton $\cX$ enters linearly in the Lagrangian, its integration yields a Dirac delta of the gauge curvature, $\delta(\cF)$. The latter implies that the remaining integration is performed over flat connections. 

Witten's result has been analyzed in the context of the Schwarzian theory, where the BF model is based on the noncompact group ${\rm SL}(2, \mathbb{R})$ \cite{Saad:2019lba}.  Here, we examine some of the most important points of this discussion, which can be summarized by the following expression of the partition function
\begin{equation}
Z=\int  \dD[\cA_{\rm flat}] \sqrt{{\rm det}(\omega)} \times \exp(-I_{\rm bndy})\,,\ 
\end{equation}
where $I_{\rm bndy}$ is boundary action presented in section \eqref{BoundaryAction}.  Furthermore, the integration measure is given by the determinant of the symplectic form $\omega$ over the space of flat connections, whose component is\footnote{The full symplectic form is obtained after anti-symmetrizing the above expression and multiplying it by its tangent basis vectors.}
\be
\omega( \delta_1\cA, \delta_2\cA)= \gamma \int{\rm tr} \left(\delta_1\cA \wedge \delta_2\cA \right)\,,
\ee
where $\delta_i\cA$ are tangent vectors on the space of flat connections $\cA=\cU^{-1} d \cU$, with $\cU$ a ${\rm SL}(2, \mathbb{R})$ group element and $\gamma$ is a constant that will be fixed in section \eqref{sym-bt} to connect with the symplectic structure on the coadjoint orbit of the Virasoro group. The subindices $i=1,2$ represent different configurations.

A more tractable expression for $\omega$ can be found after expressing $\delta_i \cA$ in terms of the one-forms $\Theta_i=\cU^{-1}\delta_i \cU$. This yields  an infinitesimal gauge transformation $\delta_i \cA=  d \Theta_i+[\cA, \Theta_i]$. By using this expression for the variation of the connection, the symplectic form becomes an integral over the thermal cycle 
\be
\omega( \delta_1\cA, \delta_2\cA)=\gamma\int dy \,{\rm tr} \left(\Theta_1 ( \d_y  \Theta_2 + [\cA_y, \Theta_2]) \right)\,.
\ee
Since the symplectic form is entirely determined by the boundary data, the function $\Theta$ becomes the canonical mode responsible for the asymptotic dynamics. Furthermore, as it was discussed in section \ref{asyIS}, transformations preserving boundary conditions are given by the value of the dilaton field
\eqref{dilatongauge}. Hence, the dynamics of $\Theta$ is controlled by the dilaton itself at the boundary
\be
\Theta_i=b^{-1}(\rho) \, x_i\, b(\rho)\,, 
\ee
where  $x_i\equiv x(\varepsilon_i,\cL;\lambda)$ and we have used the parametrization in \eqref{radial gaugetrans}.  The symplectic form can be further reduced, yielding 
\be
\omega( \delta_1\cA, \delta_2\cA)=\tfrac{2\pi \gamma}{k}\int dy \, \varepsilon_1 \delta_2 \cL=\tfrac{\gamma}{2}\int^{\beta}_0 dy \, \left[ \varepsilon_1'\varepsilon_2'' +\tfrac{4\pi}{k} (\varepsilon_1 \varepsilon_2'-\varepsilon_2 \varepsilon_1') \cL \right]\,.
\ee
or, on a coordinate-independent basis, we find
\be
\label{omegaeps}
\omega=\tfrac{\gamma}{2}\int^{\beta}_0 dy \, \left[ d\varepsilon' d\varepsilon'' + \tfrac{8\pi}{k}\cL\, d\varepsilon d\varepsilon' \right]\,.
\ee
This symplectic form is the infinitesimal limit of \eqref{symplectic virasoro} obtained in section \ref{sym-bt}.
Indeed, by performing the transformation $f(y)=y+\varepsilon(y)$ one lands on \eqref{omegaeps}, which is strictly equivalent after setting $\gamma=-k/4\pi$. 

\subsection{Quadratic contribution and entropy}
We evaluate the partition function 
\begin{equation}
\label{Zpathintegral}
Z(\beta)=\int_{{\rm SL}(2,\mathbb{R})} \dD[f] \, \sqrt{\det(\omega)}  \exp(-I_{\rm bndy}[f])\,,\\
\end{equation}
considering the infinitesimal diffeomorphism $f(y)=y+\varepsilon(y)$ and taking up to quadratic contributions of $\varepsilon$ in the boundary action. 
In the functional integral, we have made explicit that the domain of integration corresponds to the Virasoro coadjoint orbit consisting of reparametrization fields $f$ that does not belong to the little group ${\rm SL}(2,\mathbb{R})$. This corresponds to the single cover of the hyperbolic space, whose representative is $\cL_0=-\tfrac{\pi^2}{\beta^2}$.

As shown in section \ref{BoundaryAction}, we recall that the boundary action is given by the $k$-th kdV Hamiltonian 
\be
\label{geu}
I_{\rm bndy}[f]= g_k \int \, dy  \, \cH_k[f] \,.
\ee
We will compute the second order correction of \eqref{Zpathintegral} choosing the $k$-th Hamiltonian of the KdV hierachy, which can been written in terms of the reparametrization mode $f$ following the discussion in subsection \ref{Metric formalism}, that led to the equation \eqref{rep}. The boundary conditions for $f$ are
\be 
f(y+\beta)=f(y)+\beta\,,
\label{bc}
\ee
so that we use a mode expansion consistent with them,
\be
\label{me}
f(y)= y + \varepsilon(y)\,, \quad  \varepsilon(y)= \frac{\beta}{2\pi}\sum_{n} (\varepsilon^{R}_{n}+i\varepsilon^{I}_{n}) e^{i\tfrac{2\pi}{\beta} n y}\,.
\ee
The two above equations \eqref{bc} and \eqref{me} impose a reality condition on $\varepsilon$, where we have that $\varepsilon^I_{-n}=-\varepsilon^I_n$ and $\varepsilon^R_{-n}=\varepsilon^R_n$. 
Expanding the action \eqref{geu} in  $\varepsilon$, we find 
\be
 \int  dy \; \cH_{k}= \frac{ \beta \alpha_{k} (-1)^{k+1}}{ 2(2k-1)} \cL_0^k+\frac{(-1)^{k+1}}{4}\sum^{k}_{j=1} \alpha_{k-j} \cL_0^{k-j}  \int   dy \;  \varepsilon'  \cD^j_0 \varepsilon' + O(\varepsilon^3) \,,
\ee
which is very compact form to express the expanded action up quadratic contributions on $\varepsilon$ for all values of $k$, that is inhereted from the recurrence \eqref{Gerlfand2}. Here 
$$\cD_0 = 4\cL_0-\d_{y}^2\quad\text{and}\quad\alpha_{k}=\frac{(2k)!}{(k!)^2}.$$ Furthermore, the symplectic form \eqref{omegaeps} in terms of the variables $\varepsilon^R_n$ and $\varepsilon^I_n$ becomes
\be
\omega=4\pi \gamma \sum_{n}(n^3-n) d\varepsilon^{R}_n \, d\varepsilon^{I}_n\,.
\ee
Evaluating the determinant of the symplectic form and expressing the quadratic contribution in Fourier modes, the partition function associated to the $k-$th Hamiltonian \eqref{Zpathintegral} reduces to 
\be
\label{Gaussian}
Z_k(\beta)=e^{\tfrac{g_k \pi^{2k} \alpha_{k}}{ 2(2k-1)} \beta^{1-2k}} \prod_{n \geq 2} 4\pi \gamma \, (n^3- n)\, 
\int\, d\varepsilon^{R}_n \, d\varepsilon^{I}_n \,\exp\left[- 2\pi^{2k} g_k \beta^{1-2k}  \dA_{n,k} |\varepsilon_n|^2\right]\,,
\ee
where $\dA_{k,n}$ is simply a numeric factor 
\be
\dA_{n,k}=\sum^k_{j=1}(-4)^{j-1}\alpha_{k-j} n^2(n^2-1)^j \,.
\ee
Here we are integrating ruling out the zero modes, which correspond to linearized SL$(2,\mathbb{R})$ transformations of the classical solution\cite{Maldacena:2016upp}.

To ensure convergence of the above Gaussian integral, one needs to check that $\dA_{k,n}$ is positive definite. We verify that this occurs for odd values of the integer $k$ when $\beta>0$.  It is important to mention that this is a branch connected with the Schwarzian action for $k=1$. Therefore, our results can be regarded as higher derivative extension of the known case. For even values of $k$ one should consider $\beta>0$. Negative and positive values of $\beta$ are allowed by the expression \eqref{temperature}.

Performing the Gaussian integrals in \eqref{Gaussian}, one gets
\be
Z_k(\beta)=\exp\left[\tfrac{g_k \pi^{2k} \alpha_{k}}{ 2(2k-1)\beta^{2k-1}} \right]\prod_{n \geq 2} \frac{\beta^{2k-1}}{g_k (2\pi)^{2k-2} n}  \frac{\gamma}{\dN_{n,k}} \,, 
\ee
with $\dN_{n,k}= \sum^s_{j=1} 4^{j-2} \alpha_{k-j} (1-n^2)^{j-1}$. One can regularize the above infinite product by computing the average energy, 
\be
\label{jaja}
\langle E \rangle \equiv -\frac{\d}{\d \beta} \log Z_k(\beta)\,.
\ee
Interestingly enough, the factor proportional to $\dN_{n,k}$ drops out from the energy as it does not have any temperature dependence, yielding
\be
\langle E_{k} \rangle = \frac{g_k \pi^{2k} \alpha_{k}}{ 2\beta^{2k}}+\frac{2k-1}{\beta}\sum_{n\geq 2} 1=\frac{g_k \pi^{2k} \alpha_{k}}{ 2\beta^{2k}}-\frac{3(2k-1)}{2\beta}\,,
\ee
that has been regularized using the zeta function, $\zeta(0)=-\tfrac{1}{2}$. One can integrate the expression \eqref{jaja}, finding that the regularized one-loop corrected partition function gives
\be
Z^{\rm reg}_k(\beta)=\cZ_0 \exp\left[-\tfrac{g_k \pi^{2k}\alpha_{k}}{ 2(2k-1)\beta^{2k-1} }\right] (\beta J)^{-\frac{3}{2}(2k-1)} \,.
\ee
Here $\cZ_0$ is an unknown but irrelevant prefactor, as it does not exhibit any temperature dependence and can be reabsorbed in a shift. Also, the parameter $J^{2k-1}\sim g_{k}$ is a fixed constant that has been introduced so that the partition function is dimensionless. 
Considering the value of regularized one-loop partition function, one can compute the free energy, 
\begin{equation}
   -\beta F_{k}\equiv \log Z^{\rm reg}_{k}(\beta)= - \frac{g_k\pi^{2k}\alpha_{k}}{ 2(2k-1)\beta^{2k-1}}-\frac{3}{2}(2k-1)\log(\beta J)+S_{0}
\end{equation}
where $S_{0}$ is a constant associated to $\mathcal{Z}_{0}$. The entropy of the one-dimensional model in the canonical ensemble is given by
\be
\label{ST}
S=\left(1-\beta \frac{\d}{\d\beta}\right)\log Z^{\rm reg}_k(\beta)= \frac{k \alpha_k \pi^{2k}}{2k-1}\,  g_k T^{2k-1} + \frac{3}{2}(2k-1) \log (T/J)+ \dots\,,
\ee
where the dots are subleading terms in $T$. We have recovered the known result for $k=1$ \cite{Saad:2019lba, Mertens:2022irh}, where the entropy has a linear behavior on the temperature and also introduce an interesting generalization with a power law dependence. The first term is related to the classical contribution while the second is a quantum effect. In black hole physics, the Schwarzian modes captures the deviations of a near horizon extremal black holes, with a similar role in charged \cite{Almheiri:2016fws,Nayak:2018qej} and Kerr extreme perturbations \cite{Kapec:2019hro,Castro:2019crn}. Considering that near horizon geometry of these type of black holes have smaller temperature while closer to extremality, it implies quantum gravity effects become dominant in that scenario, which is still true independently of the value of $k$. On the other hand, the label $k$ becomes relevant in the classical contribution, where extremal configurations with the same value of the temperature but non-linearly perturbed with a generalized KdV dynamics, have a larger entropy, as they are associated to higher order boundary actions. 

\section{Overview and discussion}

This work proposed new boundary conditions for JT gravity on the metric and dilaton fields, that lead to a novel set of boundary actions for the effective theory. They include previous results and also generalize them to a broader phase space, dynamic equations and thermodynamic features. They keep making manifest the symmetry breaking from
$$\text{Diff}(S^{1})\rightarrow SL(2,\mathbb{R}),$$ but the deformations away AdS$_{2}$ are now generalized to infinite possible non-linear equations related with the stationary KdV hierarchy.
We would like to note that this enhancement on the asymptotic symmetries are not precisely manifest through an enlargement of them, but in the way these symmetries are realized. Indeed, they permit to construct new M\"obious invariants or higher-Schwarzian actions, rendering the same symmetry breaking previously observed, but with an enriched set of asymptotic dynamics. This work also contributes to the discussion about the construction of higher Schwarzians and their relation with the Virasoro symmetry and KdV \cite{Matone:1993tj}.
 We provide a geometric origin to our boundary conditions on the metric formalism.
   In terms of the AdS$_{2}$ metric, the seudo-Nambu Goldstone modes associated to $f$ are deviations to the geometry that could be understood as a large gauge transformation \eqref{metric}. 
 We also observed an important modification on the thermodynamics features. In the Schwarzian theory ($k=1$), the leading behavior of the entropy matches with the entropy of a nearly extremal black holes, whose effective dynamics is captured by the JT gravity. In the present case, the equation \eqref{ST} exhibits a power-law behavior. Being JT a dual theory to SYK, one may wonder whether there is a fermionic model with random interactions describing those degrees of freedom in a particular regime. An interesting proximity to our power law has appeared in the low-energy description of a supersymmetric modifications of the SYK and matrix models \cite{Lin:2013jra,Biggs:2023mfn}, where a similar scaling is obtained. It would be interesting to describe those degrees of freedom through a suitable supersymmetric modification to dilaton gravity theory. A strong hint may come from the study of the $\cN=1$ and $\cN = 2$  super Virasoro coadjoint orbits and its connection with the Lax form of the nonlinear differential equations of KdV type \cite{Bakas:1988mq}.

Other aspects to pursue has to do with the integrability conditions imposed by the well-posedness of the action principle. In this case, one could construct a boundary action composed with a linear combination of $H_{k}$, 
\be
\label{geu2}
I_{\rm bndy}[f]= \sum^{N}_{k=1}\int \, dy  \,  g_k \cH_k[f] \,.
\ee
Since the Hamiltonians $\cH_k [f]$ commute with each other, one can study more general ensembles with $g_s$ understood as chemical potentials, using generalized Gibbs ensembles. The consequences of the KdV symmetries for the relationship between individual microstates and thermodynamic ensembles has already been explored in \cite{Maloney:2018yrz,Dymarsky:2018lhf,Brehm:2019fyy}. It would be interesting to further understand how this picture is modified in the context of JT/SYK.

One remarkable aspect of the Schwarzian action is that its partition function localizes to the critical points of $H$ \cite{Stanford:2017thb}. The proof follows from the Duistermaat-Heckman formula \cite{Duistermaat:1982vw}, which states that a $U(1)$ invariant Hamiltonian integrated over a symplectic manifold is one-loop exact. To prove one-loop exactness, Stanford and Witten showed that there is a hidden fermionic symmetry in the partition function. In the case of discrete integrable models, Karki and Niemi \cite{Karki:1993bw} showed that this symmetry is precisely given by their bi-Hamiltonian structure, which give us a clue on how to extend localization theorems to the KdV hierachy exploiting its integrable properties \cite{olver1986applications}.

\section*{Acknowledgments}

This work was partially funded by Agencia Nacional de Investigaci\'on y Desarrollo (ANID) through Anillo Grant ACT210100. The author thanks Francisco Correa, Anatoly Dymarski and Alfredo P\'erez for discussions. Special gratitude to Hern\'an A. Gonz\'alez, for his collaboration in an initial stage of this work and for fruitful subsequent discussions.
\appendix  

\section{Construction of the KdV asymptotic conditions}\label{Appendix }

We start from the boundary conditions $a_{y}=- \frac{2\pi}{k}\cL(y)L_{-1}+L_{1}-2\lambda L_{0}$, that can be considered as a single SL$(2,\mathbb{R})$ sector of the Brown-Henneaux boundary conditions \cite{Brown:1986nw}, deformed by the parameter $\lambda$ that is a constant without variation. For the dilaton, we propose a fully general behaviour, 
\begin{equation}\label{boundarycond}
x= B L_{1}-C L_{-1}-2\,A L_{0},
\end{equation}
where $A,B$ and $C$ are functions of $y$.  
Considering the above, the asymptotic dynamics \eqref{eom} reduce to,
\begin{eqnarray}
A'+C-\frac{2\pi}{k}B\,\cL(y)&=&0,\\
B'+2\lambda B-2 A&=&0,\\
C'-2\lambda C+\frac{4\pi}{k}A\,\cL(y)&=&0.
\end{eqnarray}
After some manipulations, these equations lead to one equation for the function $B=\mu$, where
 $$A=\frac{1}{2} \left(\mu'(y)+2 \lambda  \mu \right),$$ and $$C=-\frac{\mu''(y)}{2}-\lambda  \mu'(y)+\frac{2 \pi  \mu \mathcal{L}(y)}{k}.$$
The adjoint scalar equation reduce to \eqref{eomdilaton} that is the generalized asymptotic dynamics that the permits to construct the KdV hierarchy. The final asymptotic form of the dilaton is then univocally determined by the choice of the gravitational field.

 One can also note that our boundary conditions \eqref{eq:kdvbc} fall in the ones proposed in \cite{Grumiller:2017qao}, although fixing the field along the generator $L_{0}$ permits to end up in a different result that provides a recursive way to originate the integrability conditions \eqref{recurrence}.
Indeed, this relaxation of the conformal boundary conditions leads to different boundary dynamics depending on the choice of the arbitrary integer $N$ in the asymptotic sum \eqref{expansion}. We show how this mechanism acts giving some examples.

Let us take $N=0$ that renders the simplest boundary conditions, where $\mu_{1}=cte$ with the associated Hamiltonian $H_{1}$ following \eqref{Ib}. Using this result on the asymptotic dynamics \eqref{eomKDV}, one also finds $\cL'=0$, that is the stationary part of the chiral equation. 

For $N=1$, the recurrence \eqref{recurrence} solves the following expression for the Lagrange multiplier,
\begin{equation}
\mu_2=-\frac{\pi}{k}\mu_{1} \cL + \bar{\mu}_2,
\end{equation}
where $\mu_{1},\bar{\mu}_2$ are constants associated to the integration of the recurrence relation. 
The dilaton equation \eqref{eomKDV} reduce to
$$3\cL\cL'-\frac{k}{\pi}\cL'''-\frac{k\, \bar{\mu}_2}{\pi\,\mu_{1}}\cL'=0.$$
Here we see that the result is a superposition of the previous field equation for $N=0$ and the stationary part of the KdV equation. One can see that chiral part can be removed by performing the field redefinition $\cL \to \cL +\frac{k\, \bar{\mu}_2}{3\pi\,\mu_{1}} $. 

 We give the last example, summing up to $N=2$ in \eqref{recurrence}. One finds
$$\mu_3=\mu_{1}\left(\frac{3 \pi ^2 \mathcal{L}^2}{2 k^2}-\frac{\pi \mathcal{L}''(y)}{k}\right)-\bar{\mu}_{2}\frac{ \pi \mathcal{L}(y)}{k}+\bar{\mu}_{3},$$
where $\bar{\mu}_{3}$ is a new integration constant. The above leads to a more complex boundary dynamics for \eqref{eomKDV},
$$\frac{\mu_{1}}{16}\left(\frac{120 \pi ^2 \mathcal{L}^2 \mathcal{L}'}{k^2}-\frac{20 \pi  \mathcal{L}''' \mathcal{L}}{k}-\frac{40 \pi  \mathcal{L}' \mathcal{L}''}{k}+\mathcal{L}^{(5)}\right)+\frac{\bar{\mu}_{2}}{4} \left(\mathcal{L}'''-\frac{3 \pi  \mathcal{L} \mathcal{L}'}{k}\right)+\bar{\mu}_{3} \mathcal{L}'=0$$
Similarly to the previous case, one can observe the KdV and the chiral parts supported by the integration constants $\bar{\mu}_{2}$ and $\bar{\mu}_{3}$. These contributions to the equation can be removed making the redefinition  $\cL \to \cL +K$,  where $K=\frac{k \bar{\mu}_2}{5 \pi  \mu_1}$ and $\bar{\mu}_3=\frac{3\bar{\mu}_{2}^{2}}{10\mu_1}$. The integration constants $\bar{\mu}_{2}$ and $\bar{\mu}_{3}$ can be also set to zero. In general, this last criteria is the one taken to generate each Hamiltonian separately evaluating $\mu_{k}$ to obtain $H_{k}$ in the form presented in \eqref{KdvHamiltonians}, with an appropriate choice of the integration constant $\mu_{1}$ to recover the multiplicative factor.



\begin{thebibliography}{99}
\bibitem{Jackiw:1984je}
R.~Jackiw,
``Lower Dimensional Gravity,''
Nucl. Phys. B \textbf{252} (1985), 343-356
doi:10.1016/0550-3213(85)90448-1.

\bibitem{Teitelboim:1983ux}
C.~Teitelboim,
``Gravitation and Hamiltonian Structure in Two Space-Time Dimensions,''
Phys. Lett. B \textbf{126} (1983), 41-45
doi:10.1016/0370-2693(83)90012-6

\bibitem{Kitaev:2015}
A.~Kitaev.
``A simple model of quantum holography.”  KITP strings seminars,
  April/May.
  
\bibitem{Sachdev_1993}
S.~Sachdev and J.~Ye,
``Gapless spin fluid ground state in a random, quantum Heisenberg magnet,''
Phys. Rev. Lett. \textbf{70} (1993), 3339
doi:10.1103/PhysRevLett.70.3339
[arXiv:cond-mat/9212030 [cond-mat]].

\bibitem{Sachdev:2010um}
S.~Sachdev,
``Holographic metals and the fractionalized Fermi liquid,''
Phys. Rev. Lett. \textbf{105} (2010), 151602
doi:10.1103/PhysRevLett.105.151602
[arXiv:1006.3794 [hep-th]].

\bibitem{Engelsoy:2016xyb}
J.~Engels\"oy, T.~G.~Mertens and H.~Verlinde,
``An investigation of AdS$_{2}$ backreaction and holography,''
JHEP \textbf{07} (2016), 139
doi:10.1007/JHEP07(2016)139
[arXiv:1606.03438 [hep-th]].

\bibitem{Jensen:2016pah} K.~Jensen,
``Chaos in AdS$_2$ Holography,''
Phys. Rev. Lett. \textbf{117} (2016) no.11, 111601
doi:10.1103/PhysRevLett.117.111601
[arXiv:1605.06098 [hep-th]].
\bibitem{Maldacena:2016upp}
J.~Maldacena, D.~Stanford and Z.~Yang,
``Conformal symmetry and its breaking in two dimensional Nearly Anti-de-Sitter space,''
PTEP \textbf{2016} (2016) no.12, 12C104
doi:10.1093/ptep/ptw124
[arXiv:1606.01857 [hep-th]].

\bibitem{Kitaev:2017awl}
A.~Kitaev and S.~J.~Suh,
``The soft mode in the Sachdev-Ye-Kitaev model and its gravity dual,''
JHEP \textbf{05} (2018), 183
doi:10.1007/JHEP05(2018)183
[arXiv:1711.08467 [hep-th]].


\bibitem{Grumiller:2015vaa}
D.~Grumiller, J.~Salzer and D.~Vassilevich,
``AdS$_{2}$ holography is (non-)trivial for (non-)constant dilaton,''
JHEP \textbf{12} (2015), 015
doi:10.1007/JHEP12(2015)015
[arXiv:1509.08486 [hep-th]].

\bibitem{Grumiller:2017qao}
D.~Grumiller, R.~McNees, J.~Salzer, C.~Valc\'arcel and D.~Vassilevich,
``Menagerie of AdS$_{2}$ boundary conditions,''
JHEP \textbf{10} (2017), 203
doi:10.1007/JHEP10(2017)203
[arXiv:1708.08471 [hep-th]].

\bibitem{Godet:2020xpk}
V.~Godet and C.~Marteau,
``New boundary conditions for AdS$_{2}$,''
JHEP \textbf{12} (2020), 020
doi:10.1007/JHEP12(2020)020
[arXiv:2005.08999 [hep-th]].

\bibitem{hotta1998asymptotic}
M.~Hotta,
``Asymptotic isometry and two-dimensional anti-de Sitter gravity,''
[arXiv:gr-qc/9809035 [gr-qc]].

\bibitem{Cadoni_1999}
M.~Cadoni and S.~Mignemi,
``Asymptotic symmetries of AdS(2) and conformal group in $d = 1$,''
Nucl. Phys. B \textbf{557} (1999), 165-180
doi:10.1016/S0550-3213(99)00398-3
[arXiv:hep-th/9902040 [hep-th]].

\bibitem{Navarro-Salas:1999zer}
J.~Navarro-Salas and P.~Navarro,
``AdS(2) / CFT(1) correspondence and near extremal black hole entropy,''
Nucl. Phys. B \textbf{579} (2000), 250-266
doi:10.1016/S0550-3213(00)00165-6
[arXiv:hep-th/9910076 [hep-th]].

\bibitem{Fu:2016vas} W.~Fu, D.~Gaiotto, J.~Maldacena and S.~Sachdev,
``Supersymmetric Sachdev-Ye-Kitaev models,''
Phys. Rev. D \textbf{95} (2017) no.2, 026009
doi:10.1103/PhysRevD.95.026009
[arXiv:1610.08917 [hep-th]].

\bibitem{Murugan:2017}
J.~Murugan, D.~Stanford and E.~Witten,
``More on Supersymmetric and 2d Analogs of the SYK Model,''
JHEP \textbf{08} (2017), 146
doi:10.1007/JHEP08(2017)146
[arXiv:1706.05362 [hep-th]].

\bibitem{Bulycheva:2018qcp}
K.~Bulycheva,
``$ \mathcal{N}=2 $ SYK model in the superspace formalism,''
JHEP \textbf{04} (2018), 036
doi:10.1007/JHEP04(2018)036
[arXiv:1801.09006 [hep-th]].

\bibitem{Biggs:2023mfn}
A.~Biggs, J.~Maldacena and V.~Narovlansky,
``A supersymmetric SYK model with a curious low energy behavior,''
[arXiv:2309.08818 [hep-th]].



\bibitem{Bulycheva_2017}
K.~Bulycheva,
``A note on the SYK model with complex fermions,''
JHEP \textbf{12} (2017), 069
doi:10.1007/JHEP12(2017)069
[arXiv:1706.07411 [hep-th]].

\bibitem{Chaturvedi:2018uov}
P.~Chaturvedi, Y.~Gu, W.~Song and B.~Yu,
``A note on the complex SYK model and warped CFTs,''
JHEP \textbf{12} (2018), 101
doi:10.1007/JHEP12(2018)101
[arXiv:1808.08062 [hep-th]].
\bibitem{Gu_2020}
Y.~Gu, A.~Kitaev, S.~Sachdev and G.~Tarnopolsky,
``Notes on the complex Sachdev-Ye-Kitaev model,''
JHEP \textbf{02} (2020), 157
doi:10.1007/JHEP02(2020)157
[arXiv:1910.14099 [hep-th]].

\bibitem{Almheiri:2014cka}
A.~Almheiri and J.~Polchinski,
``Models of AdS$_{2}$ backreaction and holography,''
JHEP \textbf{11} (2015), 014
doi:10.1007/JHEP11(2015)014
[arXiv:1402.6334 [hep-th]].

\bibitem{aharonov1969necessary}
Harmelin,~R. ``Aharonov invariants and univalent functions,'' Israel J. Math.  \textbf{43}, 244–254 (1982). https://doi.org/10.1007/BF02761945

\bibitem{Tamanoi1996HigherSO}
Tamanoi,~ H. ``Higher Schwarzian operators and combinatorics of the Schwarzian derivative," Math. Ann. \textbf{305}, 127–151 (1996). https://doi.org/10.1007/BF01444214


\bibitem{kim2009some}
Kim,~ Seong-A. (2009). ``Some generalized higher Schwarzian operators, " The Pure and Applied Mathematics,  \textbf{16},  (2009)

\bibitem{Galajinsky:2023btq}
A.~Galajinsky,
``Remarks on higher Schwarzians,''
Phys. Lett. B \textbf{843} (2023), 138042
doi:10.1016/j.physletb.2023.138042
[arXiv:2302.00317 [hep-th]].

\bibitem{Krivonos:2024jpo}
S.~Krivonos,
``Origin of higher Schwarzians,''
Phys. Rev. D \textbf{109} (2024) no.6, 065029
doi:10.1103/PhysRevD.109.065029
[arXiv:2401.10532 [hep-th]].

\bibitem{Lin:2013jra}
Y.~H.~Lin, S.~H.~Shao, Y.~Wang and X.~Yin,
``A Low Temperature Expansion for Matrix Quantum Mechanics,''
JHEP \textbf{05} (2015), 136
doi:10.1007/JHEP05(2015)136
[arXiv:1304.1593 [hep-th]].

\bibitem{Okuyama:2019xbv}
K.~Okuyama and K.~Sakai,
``JT gravity, KdV equations and macroscopic loop operators,''
JHEP \textbf{01} (2020), 156
doi:10.1007/JHEP01(2020)156
[arXiv:1911.01659 [hep-th]].


\bibitem{Blommaert:2022lbh}
A.~Blommaert, J.~Kruthoff and S.~Yao,
``An integrable road to a perturbative plateau,''
JHEP \textbf{04} (2023), 048
doi:10.1007/JHEP04(2023)048
[arXiv:2208.13795 [hep-th]].

\bibitem{Sasaki:1987mm}
R.~Sasaki and I.~Yamanaka,
``Virasoro Algebra, Vertex Operators, Quantum {Sine-Gordon} and Solvable Quantum Field Theories,''
Adv. Stud. Pure Math. \textbf{16} (1988), 271-296
RRK-87-3.

\bibitem{Kupershmidt:1989bf}
B.~A.~Kupershmidt and P.~Mathieu,
``Quantum Korteweg-de Vries Like Equations and Perturbed Conformal Field Theories,''
Phys. Lett. B \textbf{227} (1989), 245-250
doi:10.1016/S0370-2693(89)80030-9

\bibitem{Bazhanov:1994ft}
V.~V.~Bazhanov, S.~L.~Lukyanov and A.~B.~Zamolodchikov,
``Integrable structure of conformal field theory, quantum KdV theory and thermodynamic Bethe ansatz,''
Commun. Math. Phys. \textbf{177} (1996), 381-398
doi:10.1007/BF02101898
[arXiv:hep-th/9412229 [hep-th]].

\bibitem{Asrat:2020jsh}
M.~Asrat,
``KdV charges and the generalized torus partition sum in $T \bar T$ deformation,''
Nucl. Phys. B \textbf{958} (2020), 115119
doi:10.1016/j.nuclphysb.2020.115119
[arXiv:2002.04824 [hep-th]].

\bibitem{Perez:2016vqo}
A.~P\'erez, D.~Tempo and R.~Troncoso,
``Boundary conditions for General Relativity on AdS$_{3}$ and the KdV hierarchy,''
JHEP \textbf{06} (2016), 103
doi:10.1007/JHEP06(2016)103
[arXiv:1605.04490 [hep-th]].

\bibitem{Gonzalez:2018jgp}
H.~A.~Gonz\'alez, J.~Matulich, M.~Pino and R.~Troncoso,
``Revisiting the asymptotic dynamics of General Relativity on AdS$_{3}$,''
JHEP \textbf{12} (2018), 115
doi:10.1007/JHEP12(2018)115
[arXiv:1809.02749 [hep-th]].

\bibitem{Cardenas:2021vwo}
M.~C\'ardenas, F.~Correa, K.~Lara and M.~Pino,
``Integrable Systems and Spacetime Dynamics,''
Phys. Rev. Lett. \textbf{127} (2021) no.16, 161601
doi:10.1103/PhysRevLett.127.161601
[arXiv:2104.09676 [hep-th]].

\bibitem{Fukuyama:1985gg}
T.~Fukuyama and K.~Kamimura,
``Gauge Theory of Two-dimensional Gravity,''
Phys. Lett. B \textbf{160} (1985), 259-262
doi:10.1016/0370-2693(85)91322-X

\bibitem{Fefferman:2007rka}
C.~Fefferman and C.~R.~Graham,
``The ambient metric,''
Ann. Math. Stud. \textbf{178} (2011), 1-128
[arXiv:0710.0919 [math.DG]].

\bibitem{Roberts:2012aq}
M.~M.~Roberts,
``Time evolution of entanglement entropy from a pulse,''
JHEP \textbf{12} (2012), 027
doi:10.1007/JHEP12(2012)027
[arXiv:1204.1982 [hep-th]].

\bibitem{Witten:1987ty}
E.~Witten,
``Coadjoint Orbits of the Virasoro Group,''
Commun. Math. Phys. \textbf{114} (1988), 1
doi:10.1007/BF01218287


\bibitem{Bakas:1988bq}
I.~Bakas,
``Orbits of Diff S1 in the space of quadratic differentials''
Nucl. Phys. B Proc. Suppl. \textbf{6} (1989), 137-139
doi:10.1016/0920-5632(89)90419-2

\bibitem{Alekseev:1988ce}
A.~Alekseev and S.~L.~Shatashvili,
``Path Integral Quantization of the Coadjoint Orbits of the Virasoro Group and 2D Gravity,''
Nucl. Phys. B \textbf{323} (1989), 719-733
doi:10.1016/0550-3213(89)90130-2

\bibitem{Babelon:2003qtg}
O.~Babelon, D.~Bernard and M.~Talon,``Introduction to Classical Integrable Systems,''
Cambridge University Press, 2003,
ISBN 978-0-521-03670-2, 978-0-511-53502-4
doi:10.1017/CBO9780511535024

\bibitem{Witten:1991we}
E.~Witten,
``On quantum gauge theories in two-dimensions,''
Commun. Math. Phys. \textbf{141} (1991), 153-209
doi:10.1007/BF02100009

\bibitem{Saad:2019lba}
P.~Saad, S.~H.~Shenker and D.~Stanford,
``JT gravity as a matrix integral,''
[arXiv:1903.11115 [hep-th]].

\bibitem{Mertens:2022irh}
T.~G.~Mertens and G.~J.~Turiaci,
``Solvable models of quantum black holes: a review on Jackiw\textendash{}Teitelboim gravity,''
Living Rev. Rel. \textbf{26} (2023) no.1, 4
doi:10.1007/s41114-023-00046-1
[arXiv:2210.10846 [hep-th]].

\bibitem{Almheiri:2016fws}
A.~Almheiri and B.~Kang,
``Conformal Symmetry Breaking and Thermodynamics of Near-Extremal Black Holes,''
JHEP \textbf{10} (2016), 052
doi:10.1007/JHEP10(2016)052
[arXiv:1606.04108 [hep-th]].

\bibitem{Nayak:2018qej}
P.~Nayak, A.~Shukla, R.~M.~Soni, S.~P.~Trivedi and V.~Vishal,
JHEP \textbf{09} (2018), 048
doi:10.1007/JHEP09(2018)048
[arXiv:1802.09547 [hep-th]].

\bibitem{Kapec:2019hro}
D.~Kapec and A.~Lupsasca,
Class. Quant. Grav. \textbf{37} (2020) no.1, 015006
doi:10.1088/1361-6382/ab519e
[arXiv:1905.11406 [hep-th]].

\bibitem{Castro:2019crn}
A.~Castro and V.~Godet,
``Breaking away from the near horizon of extreme Kerr,''
SciPost Phys. \textbf{8} (2020) no.6, 089
doi:10.21468/SciPostPhys.8.6.089
[arXiv:1906.09083 [hep-th]].

\bibitem{Matone:1993tj}
M.~Matone,
``Uniformization theory and 2-D gravity. 1. Liouville action and intersection numbers,''
Int. J. Mod. Phys. A \textbf{10} (1995), 289-336
doi:10.1142/S0217751X95000139
[arXiv:hep-th/9306150 [hep-th]].

\bibitem{Bakas:1988mq}
I.~Bakas,
``Conformal Invariance, the {KdV} Equation and Coadjoint Orbits of the Virasoro Algebra,''
Nucl. Phys. B \textbf{302} (1988), 189-203
doi:10.1016/0550-3213(88)90241-6

\bibitem{Maloney:2018yrz}
A.~Maloney, G.~S.~Ng, S.~F.~Ross and I.~Tsiares,
``Generalized Gibbs Ensemble and the Statistics of KdV Charges in 2D CFT,''
JHEP \textbf{03} (2019), 075
doi:10.1007/JHEP03(2019)075
[arXiv:1810.11054 [hep-th]].

\bibitem{Dymarsky:2018lhf}
A.~Dymarsky and K.~Pavlenko,
``Generalized Gibbs Ensemble of 2d CFTs at large central charge in the thermodynamic limit,''
JHEP \textbf{01} (2019), 098
doi:10.1007/JHEP01(2019)098
[arXiv:1810.11025 [hep-th]].

\bibitem{Brehm:2019fyy}
E.~M.~Brehm and D.~Das,
``Korteweg\textendash{}de Vries characters in large central charge CFTs,''
Phys. Rev. D \textbf{101} (2020) no.8, 086025
doi:10.1103/PhysRevD.101.086025
[arXiv:1901.10354 [hep-th]].

\bibitem{Stanford:2017thb}
D.~Stanford and E.~Witten,
``Fermionic Localization of the Schwarzian Theory,''
JHEP \textbf{10} (2017), 008
doi:10.1007/JHEP10(2017)008
[arXiv:1703.04612 [hep-th]].

\bibitem{Duistermaat:1982vw}
J.~J.~Duistermaat and G.~J.~Heckman,
``On the Variation in the cohomology of the symplectic form of the reduced phase space,''
Invent. Math. \textbf{69} (1982), 259-268
doi:10.1007/BF01399506

\bibitem{Karki:1993bw}
T.~Karki and A.~J.~Niemi,
``On the Duistermaat-Heckman formula and integrable models,''
[arXiv:hep-th/9402041 [hep-th]].


\bibitem{olver1986applications}
P.J. ~Olver, ``Applications of Lie Groups to Differential Equations,'', Advances in Physical Geochemistry, World Publishing Company, 1986.

\bibitem{Brown:1986nw}
J.~D.~Brown and M.~Henneaux,
``Central Charges in the Canonical Realization of Asymptotic Symmetries: An Example from Three-Dimensional Gravity,''
Commun. Math. Phys. \textbf{104} (1986), 207-226
doi:10.1007/BF01211590

\end{thebibliography}
\end{document}